\documentclass[revtex4]{emulateapjold}

\newcommand{\mytilde}{\raise.17ex\hbox{$\scriptstyle\sim$}}
\newcommand{\brk}[1]{$[$#1$]$}
\newcommand{\kepler}{{\it Kepler}}
\newcommand{\jwst}{{\it JWST}}
\newcommand{\hst}{{\it HST}}

\newcommand{\mps}{m~s$^{-1}$}

\newcommand{\Msun}{${M_\odot}$}
\newcommand{\Rsun}{${R_\odot}$}

\newcommand{\Mear}{${M_\oplus}$}

\newcommand{\Mstar}{${M_\star}$}
\newcommand{\Rstar}{${R_\star}$}
\newcommand{\Teff}{${T_{\rm eff}}$}
\newcommand{\Msini}{$M_{p}\, {\rm sin}i$}
\newcommand{\Tobs}{${T_{obs}}$}
\newcommand{\nobs}{${n}$}
\newcommand{\Tdur}{$\tau_{dur}$}
\newcommand{\Logg}{$\log{\rm g}$}

\newcommand{\Mearth}{{\it MEarth}}
\newcommand{\HARPS}{{\it HARPS}}
\newcommand{\bff}{}

\slugcomment{Submitted}

\begin{document}

\title{Transit and Radial Velocity Survey Efficiency Comparison for a Habitable Zone Earth}

\author{
Christopher~J.~Burke\altaffilmark{1} \&
P.~R.~McCullough\altaffilmark{2,3}
}

\email{christopher.j.burke@nasa.gov}
\altaffiltext{1}{SETI Institute/NASA Ames Research Center, Moffett Field, CA 94035}
\altaffiltext{2}{Space Telescope Science Institute, 3700 San Martin Dr., Baltimore, MD 21218}
\altaffiltext{3}{Department of Physics and Astronomy, Johns Hopkins University, 3400 North Charles Street, Baltimore, MD 21218}

\begin{abstract}

Transit and radial velocity searches are two techniques for
identifying nearby extrasolar planets to Earth that transit bright
stars.  Identifying a robust sample of these exoplanets around bright
stars for detailed atmospheric characterization is a major
observational undertaking.  In this study we describe a framework that
answers the question of whether a transit or radial velocity survey is
more efficient at finding transiting exoplanets given the same amount
of observing time.  Within the framework we show that a transit
survey's window function can be approximated using the hypergeometric
probability distribution.  We estimate the observing time required for
a transit survey to find a transiting Earth-sized exoplanet in the HZ
with an emphasis on late type stars.  We also estimate the radial
velocity precision necessary to detect the equivalent HZ Earth-mass
exoplanet that also transits when using an equal amount of observing
time as the transit survey.  We find that a radial velocity survey
with $\sigma_{rv}\sim$0.6~\mps\ precision has comparable efficiency in
terms of observing time to a transit survey with the requisite
photometric precision $\sigma_{phot}\sim$300~ppm to find a transiting
Earth-sized exoplanet in the HZ of late M~dwarfs.  For Super-Earths, a
$\sigma_{rv}\sim$2.0~\mps\ precision radial velocity survey has
comparable efficiency to a transit survey with
$\sigma_{phot}\sim$2300~ppm.

\end{abstract}

\keywords{eclipses -- methods:statistical -- planetary systems -- surveys -- techniques:photometric -- techniques:radial velocities}

\section{Introduction}\label{sec:intro}

Transiting extrasolar planets\footnote{Hereafter, ``planets'' means
  ``extrasolar planets.''}  of bright host stars provide a unique
opportunity for physical characterization in ways impossible for
non-transiting planets that are also too close to their host stars for
direct imaging.  Currently, the transit and radial velocity methods
dominate the growth in planet discoveries both overall and for planets
suitable for detailed physical
characterization\footnote{http://exoplanetarchive.ipac.caltech.edu}.

This paper addresses the question, ``of these two methods, transits or
radial velocities, which is more efficient at discovering planets that
transit bright stars?''  This is a complex question given the
astrophysical effects (e.g.\ planet mass-radius relation, stellar
mass-radius relation, stellar mass-luminosity relation, radial
velocity jitter, stellar activity, etc.) and technical challenges
(e.g.\ instrumentation, telescope aperture, available observing time,
costs of operation, available funding opportunities, etc.) that
influence the planet yield of a survey.  To compare the performance of
a radial velocity survey to that of a transit survey, we adopt a
parametric model for each.  With some simplifying assumptions, we
quantify the precision of a radial velocity survey that will achieve
the same level of search completeness as a transit survey using the
same amount of observing time.

For any transit survey, the vast majority of observations of any
particular star cannot contribute directly to the detection of a
transit, simply because they are made when the planet is not
transiting.
To be successful, a transit survey must make a sufficiently large
number of independent observations in order that the probability
$p_{\ge k}$ of witnessing $k$ or more transit(s) matches or exceeds a
design requirement.  For example, the four-year Kepler mission was
designed to witness $k \ge 3$ transits of Earth analogs with orbital
periods $\la 1$ year \citep{BOR10}.  For a specific set of observation
times of a transit survey, the {\it window function}, $p_{\ge k}(P)$,
describes the visibility of the repetitive transit signal of any given
period, i.e. the planet's orbital period, $P$ \citep{GAU00}.  A survey's
window function can be computed numerically with high fidelity, either
before or after the actual observations, by Monte Carlo simulations
\citep{BUR06,VON09}.

However, analytic approximations to the window function are quicker
and more convenient to compute than Monte Carlo simulations
\citep{DEE04,BEA08,FLE08}. In addition, an analytic approximation is
readily inverted to predict the minimum observing time necessary to
exceed a given probability $p_{\ge k}(P)$ of witnessing $k$ or more
transits of a given $P$.  We show in the appendix that the
hypergeometric distribution provides a high fidelity statistical
approximation to the window function and its derivation gives physical
insight into the design of a transit survey or interpretation of
results from them.  Most importantly we show how the transit survey
window function is primarily dependent upon the number of independent
observing epochs, which enables us to compare the performance of a
transit survey to a radial velocity survey for detecting transiting
planets.

For a radial velocity survey, the situation is quite different:
nearly every measurement may contribute to the planet detection
because the star's radial velocity can be measured at any time and
is constantly changing in response to the orbiting planet. (We write
``nearly'' every measurement because multiple observations made at
the same epoch could be redundant if one would suffice at that
particular orbital phase.) To estimate the detection sensitivity
of a radial velocity survey, we adopt the prescription outlined by
\citet{CUM04}.

{\bff Our analysis ends at the phase of a survey when the significance of
the detection reaches a suitable level such that follow-up
observations to refine the ephemeris and confirm the detection are
warranted.  Thus, we neglect budgeting for the additional
observational effort required to follow-up on a significant signal as
the effort required depends very strongly on the nature of the signal
and remaining ambiguities of the signal that must be ruled out.  In
the case of a transit survey a detection from a single to few events
leads to ambiguity in the orbital period \citep{STE12,BER13} that can
be difficult to reconcile with low signal to noise detections.  The
numerous sources of astrophysical false positives to the transiting
planet signal need to be investigated \citep{BRO03,TOR04,ODO06A} along
with statistical validation \citep{TOR11,FRE11,BAR13} or mass
measurement confirmation \citep{CHA09,QUE09,BAT11}.  In the case of a
radial velocity detection, in addition to the photometric confirmation
of a transit signal \citep{HEN00,CHA00}, typically additional radial
velocity observations are required to refine the ephemeris to a
suitable level for scheduling the photometric observations
\citep{GIL07,KAN09,WIN11}.  Also, radial velocity detections must contend
with the astrophysical false positives induced by stellar variability
\citep{SAA97,BOI11,GOM12}.}

We apply our prescription to address two representative case studies.
First, we compare our models of the performance of a particular
transiting-planet finder, \Mearth\ and its southern extension, \Mearth-South
\citep{NUT08,IRW11}, to that of a particular radial velocity
instrument, \HARPS\ \citep{BON11} and its northern twin, \HARPS-N
\citep{COS12,COV13}, for detecting a Super-Earth transiting at the
inner edge of the habitable zone (HZ) as a function of stellar mass.
In the second case, we study the relative performances of these planet
finding techniques for finding Earth-sized planets transiting at the
inner edge of the HZ of M dwarf stars.  In each case, we use the inner
edge of the HZ simply because planets there will have shorter periods
than any other planets in their respective HZs and hence will be more
easily detected by either method.

This paper is organized as follows.  In Section~\ref{sec:transit} the
transit survey window function is introduced, and in the appendix we
derive analytic approximations to it.  This leads to simple
approximations for the observing time required to reach a specified
level of completeness of a transit survey. The detection sensitivity
of a radial velocity survey is inverted in Section~\ref{sec:rv} in order to
determine the observing time required for it to reach an appropriate
level of completeness to compete with the transit survey.  The
implications of this study as applied to the detection of a transiting
Super-Earth and an Earth analog at the inner edge of the HZ in
Section~\ref{sec:disc} and its conclusions are given in
Section~\ref{sec:conclusion}.

\section{Transit Survey Performance}\label{sec:transit}

We can approach the statistical modeling of a transit survey window
function in two ways. We can ask,
\begin{quotation}
A) each time the planet transits, were we observing?
\end{quotation}
or
\begin{quotation}
B) each time we observe, is the planet transiting?
\end{quotation}
\noindent For our statistical model, we want {\it independent} trials,
so to form a trial, we group observations over an appropriate time
interval.  Previous work \citep[e.g. Eq. A2 of][]{DEE04} adopted
approach A, where each independent planetary {\it orbital period} interval
is a binomial trial with probability of our witnessing it equal to the
average fraction of time spent observing $f_{\rm cov}$.  In
approach B (see Appendix), the appropriate independent time
interval is the {\it transit duration},
\begin{equation}
\tau_{dur}=0.058\left(\frac{M_{\star}}{M_{\odot}}\right)^{-1/3}\left(\frac{R_{\star}}{R_{\odot}}\right)\left(\frac{P}{{\rm 1\, day}}\right)^{1/3} {\rm [day]}, \label{eq:tdur}
\end{equation}
for arbitrary stellar mass, $M_{\star}$, radius, $R_{\star}$, and orbital period, $P$.
We have assumed the average relative chord length, $\pi/4$ and
eccentricity $e=0$ \citep{GIL00,SEA03}. Effects of non-zero
eccentricity on transit survey yield are addressed by \citet{BAR07}
and \citet{BUR08}. Approach B has the advantage of being readily
related to individual observations parceled into lengths of the
transit duration.  The window function is determined by the number of
unique observing epochs, \nobs, of duration $\tau_{dur}$.  A set of
observation exposures at a cadence shorter than the transit duration
will improve the signal to noise ratio of the detection, but count as
a single parcel increment to \nobs\ for the purposes of the window
function.\footnote{We are explicitly separating out the dependence of
  transit detection on the window function and signal to noise ratio
  \citep{GAU00}.}  Thus, \nobs\ is not necessarily the number of
exposures (see Section~\ref{sec:hyperg}).

Regardless of approach (A or B), the physical situation is the same:
in the time span, \Tobs, between the first and last observations of a
particular star with a transiting planet of orbital period, $P$, the
expected number of transits is\footnote{There are either
  $M=\lfloor$\Tobs/P$\rfloor$ or $M=\lfloor$\Tobs/P$\rfloor+1$
  transits where $\lfloor$ is the truncate to nearest integer
  operation.  For uniform and random relative phase of the planetary
  transits with respect to the set of observation epochs, the
  expectation value of $\langle M\rangle=$\Tobs/$P$}
\begin{equation}
M = T_{\rm obs} / P. \label{eq:M}
\end{equation}
From the probability, $p_k$, of witnessing exactly $k$ transits
(derived in the appendix), the probability, $p_{\geq k}$, of
witnessing $k$ or more transits is given by
\begin{equation}
p_{\geq k}=1-\sum\limits_{i=0}^{k-1}p_k.\label{eq:probN}
\end{equation}
A transit survey may require detection of $k\geq 2$ transits in order
to know the period and hence predict future transits, or it may
require $k\geq 3$ transits in order to validate the detection, i.e.
confirm that the times separating the transits are all consistent with
a particular orbital period.

\begin{figure}
\includegraphics[trim=0.3in 0.15in 0.6in 0.4in,scale=0.5,clip=true]{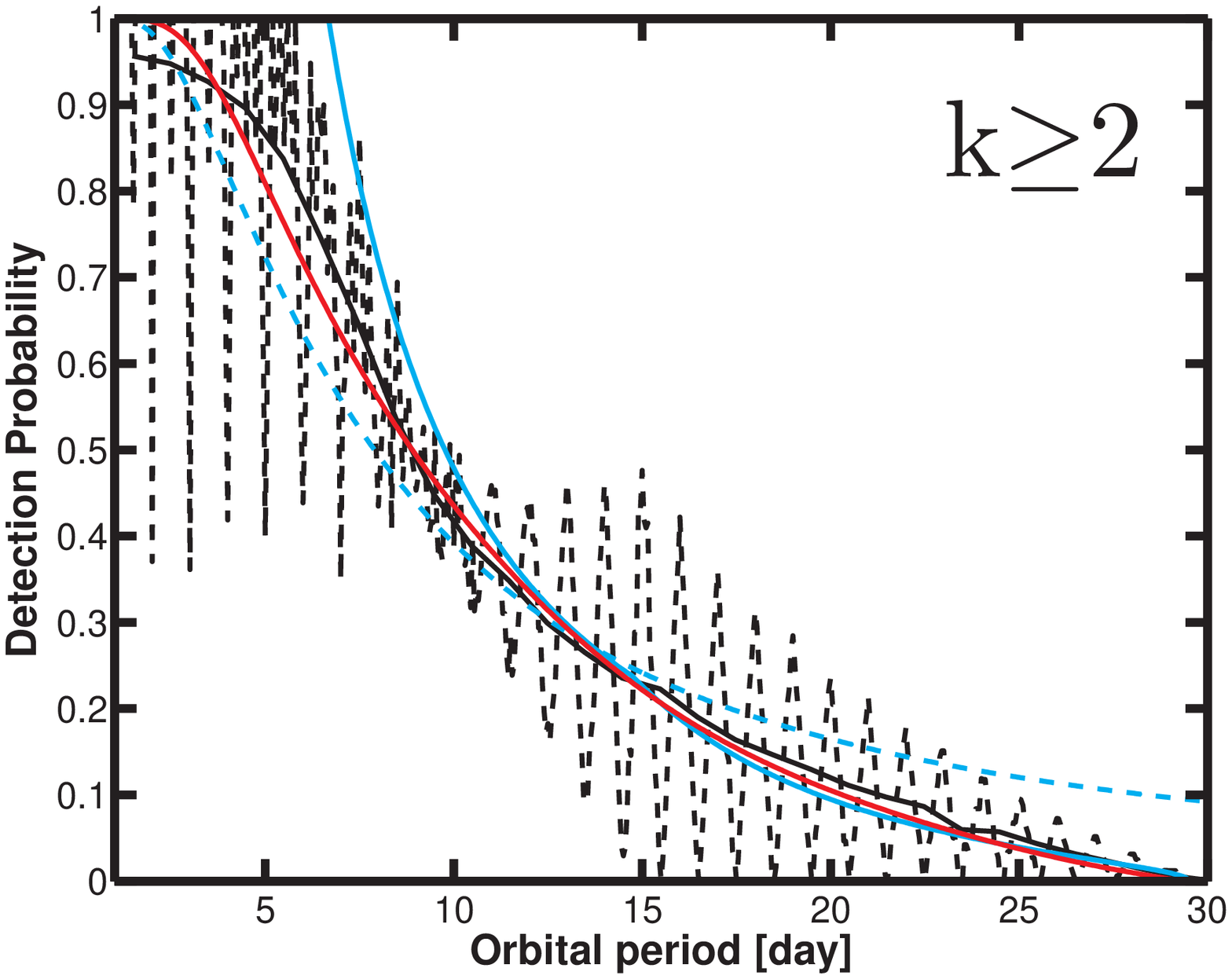}
\includegraphics[trim=0.3in 0.15in 0.6in 0.4in,scale=0.5,clip=true]{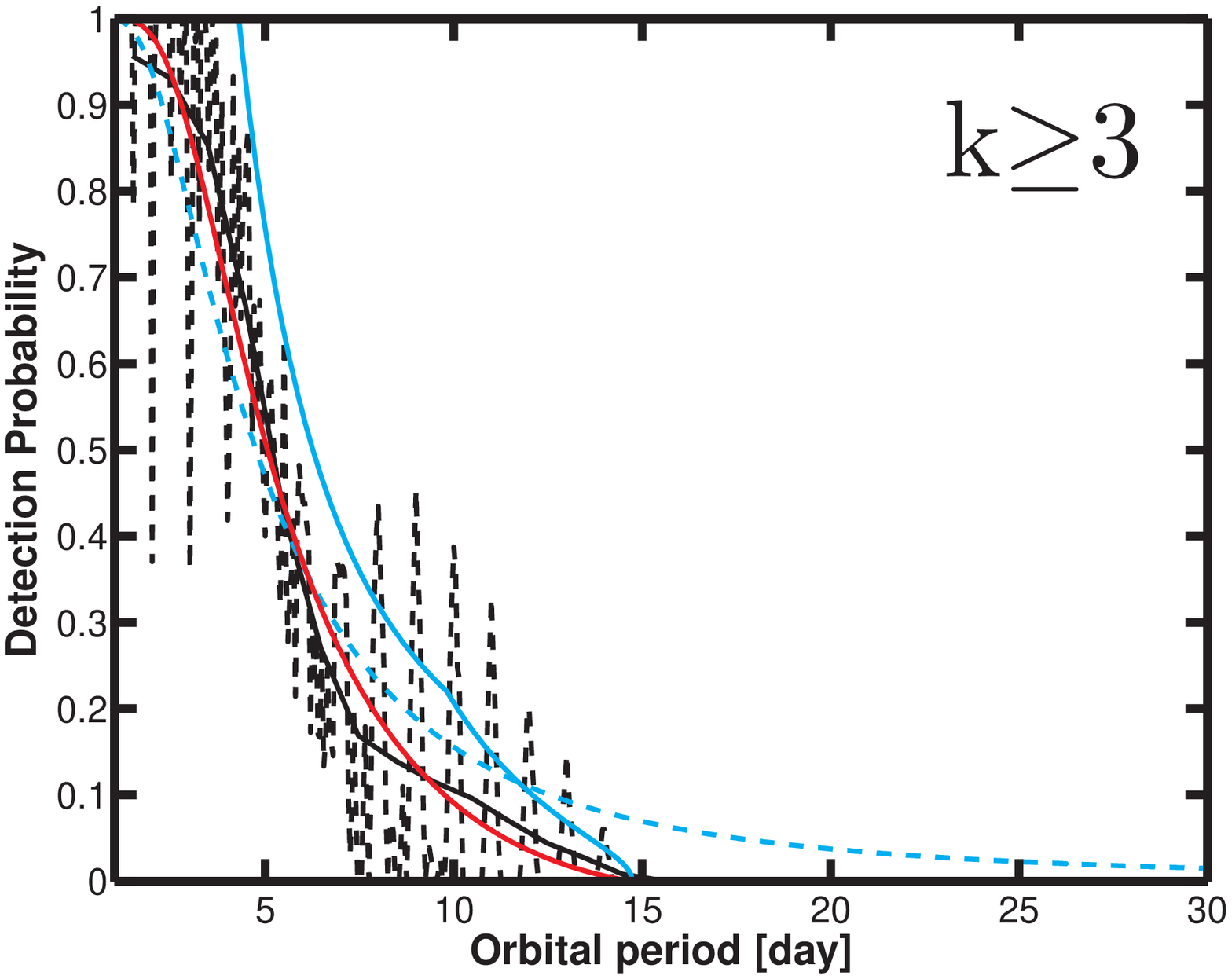}
\caption{
Left: exact numerical window function for a hypothetical ground-based
transit survey that observes for 30 consecutive days with observations
for 8 hours each night and requires two transits for detection (dashed black line).  Average of the numerical window function using a 1-day moving average (solid black line) compared to the hypergeometric model (solid red line) from this study along with its' Poisson model limit (dash blue line).  The window function approximation of \citet{BEA08} (solid blue line) is also shown.  Right: same as left panel, but requiring three transits for detection.
\label{fig:beattycomp}}
\end{figure}

In the appendix we have isolated the mathematics of approximating the
window function analytically.
Figure~\ref{fig:beattycomp} shows the
window function for a hypothetical transit survey that operates in a
mode dedicated to a single field for 30 consecutive days with
observations for 8 hours each night.  The specifications for the
hypothetical transit survey are the same as presented in Figure~1 of
\citet{BEA08} to demonstrate their analytical representation of a
transit survey and is typical for the first generation of ground-based
Hot Jupiter transit surveys ({\it HAT} -- \citet{BAK07}; {\it KELT} --
\citet{BEA12}; {\it OGLE} -- \citet{KON03}; {\it Qatar} -- \citet{ALS11};
      {\it TrES} -- \citet{ALO04}; {\it WASP} -- \citet{COL07}; {\it
        XO} -- \citet{MCC06}; and others).

\begin{figure}
\includegraphics[trim=0.3in 0.15in 0.6in 0.4in,scale=0.5,clip=true]{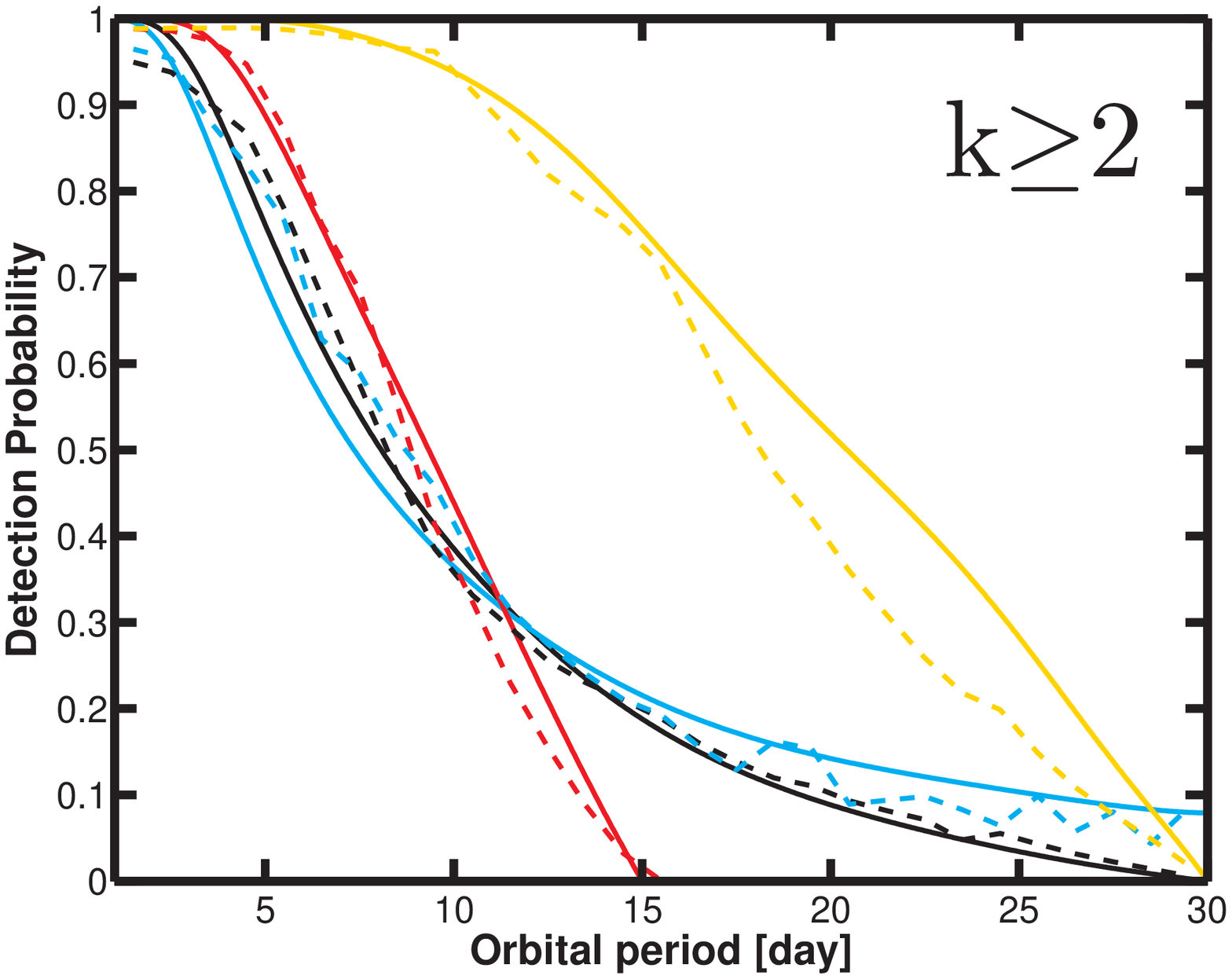}
\includegraphics[trim=0.3in 0.15in 0.6in 0.4in,scale=0.5,clip=true]{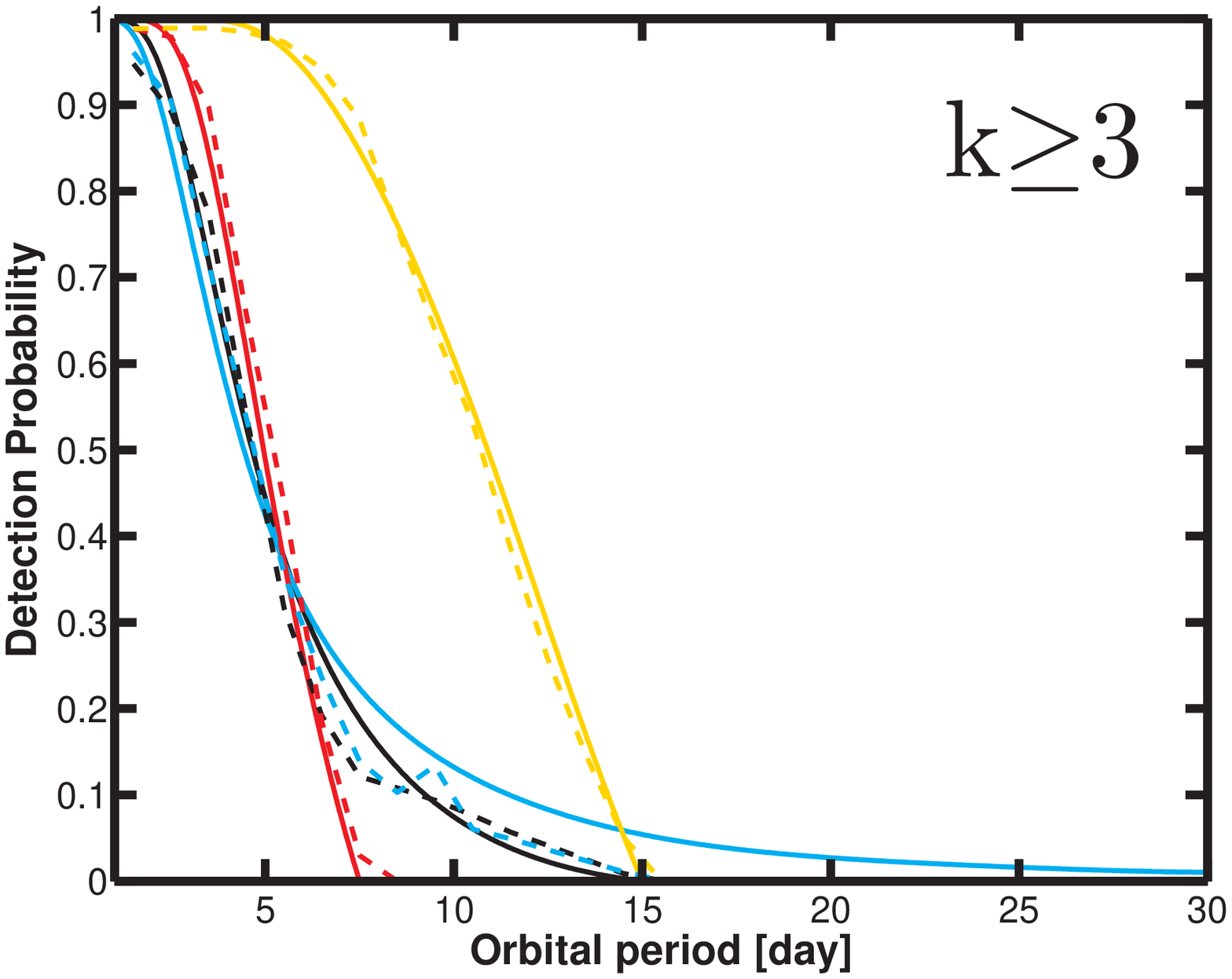}
\caption{
Demonstrating the strong dependence of a transit survey window function on $n$ compared to the weak dependence on \Tobs\ {\bff (see Section~\ref{sec:numericalcomp}).  The following three observing scenarios have the same n, but vary in \Tobs.}   We define the baseline window function for a hypothetical ground-based transit survey that observes for 7 hours each night for \Tobs=30 days using the hypergeometric model (solid black line) and 1-day moving average of exact numerical model (dashed black line).  {\bff Alternatively,} we simulate a window function for a longitudinally-distributed observing network with half, \Tobs=15 day, but increased observing coverage to equal $n$ of the baseline case (hypergeometric model, solid red line, and moving-average numerical model, dashed red line).  Finally, we simulate a seasonally-distributed transit survey window function with two observing campaigns of 15 consecutive observing days for 7 hours each night with the two campaigns separated by a year (hypergeometric model, solid blue line, and moving-average numerical model, dashed blue line).  {\bff We compare the previous scenarios to a scenario with 2$n$ from the baseline case} using \Tobs=30 day, but increased observing coverage each day (hypergeometric mode, solid yellow line, and moving-average numerical model, dashed yellow line).  Showing results requiring $k\geq2$ (left panel) and $k\geq3$ (right panel) transits.
\label{fig:alternatobs}}
\end{figure}

The window function presented in this study and others
\citep[e.g.][]{DEE04,BEA08} provide convenient approximations to the
average performance of a transit survey.  These approximations may be
useful for design purposes and for real time updates to a transit
survey in progress.  Also, this study articulates that transit survey
completeness, in terms of the window function, derives primarily from
the number of distinct observing epochs, \nobs, and only secondarily from the
time span \Tobs\ (see Figure~\ref{fig:alternatobs} {\bff and discussion in Section~\ref{sec:numericalcomp}})\footnote{In general, \nobs$\leq$ the number of exposures (see Section~\ref{sec:hyperg})}.  That is,
observations can be spread over an observing season or years rather
than concentrated in a single observing run.  A consequence is that a
longitudinally-distributed network (e.g. HAT-Net) is not necessarily
more observationally efficient at finding planets per unit of
observing time than a single site (e.g. WASP), all other factors being
equal. However, a longitudinally-distributed network of N telescopes
can retire a particular field of view of targets in a smaller number
of days than could be accomplished by the same N telescopes located at
a single site at the same latitude as the telescopes of the
network. Also, continuous observations avoid the ambiguities (e.g. of
orbital period) inherent in interpreting observations with many gaps
where transits can hide.

\begin{figure}
\includegraphics[trim=0.3in 0.15in 0.4in 0.0in,clip=true,scale=0.5]{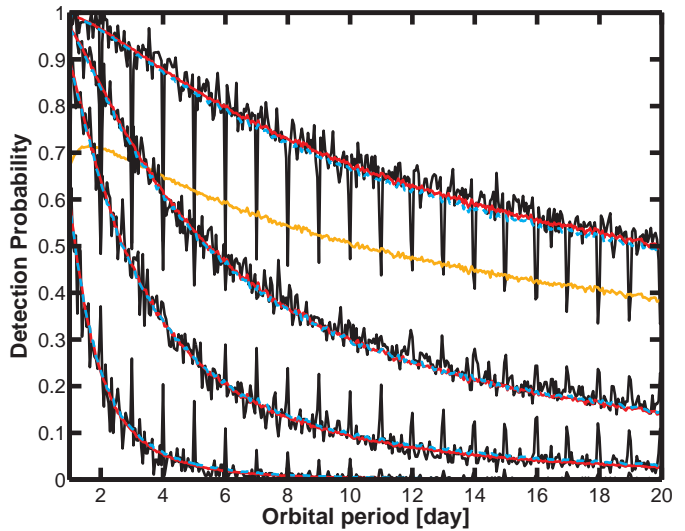}
\caption{ Exact numerical window function with \nobs=200 requiring
  observing at least one, two, three, and five transits (solid black
  lines top to bottom) compared to the hypergeometric model (solid red
  lines) and its Poisson approximation model (dash blue lines) for a typical observing time series over the course of an
  observing season.  The analytic window function employing the
  triggering on a single transit mode of operation (orange line)
  is also shown.
\label{fig:winfunc}}
\end{figure}

Figure~\ref{fig:winfunc} shows the window function for a transit
survey that operates in a mode cycling between targets such as
employed by the \Mearth\ project \citep{NUT08,BER13}.
\Mearth\ surveys the brightest M~dwarfs by cycling between targets to
obtain several observation per target per night.  The observation span
is an observing season for each target.  We simulate this mode of
operation through specifying a time series with \nobs$=200$
observations that is expected for a target from a single observatory
over the course of an observing season including losses due to
weather.  {\bff As a proxy for an observing time series we begin with a
continuous PWV time series downloaded from the NOAA Earth System
Research Laboratory Ground-Based GPS Meteorology
program\footnote{http://www.gpsmet.noaa.gov} \citep{WOL00} for $~$9
months starting September, 1 2008 with a 30 minute cadence from
Flagstaff, AZ.  We use the celestial coordinates RA 7h Dec +30,
optimized for the middle of the observing season and observatory
latitude, to find the airmass for each PWV observations.  Only values
when the airmass $<$ 1.8, PWV $<$ 10 mm (to crudely simulate when the
weather conditions would be too poor for observations), and Sun
altitude $<$ -5 deg are used to simulate times when observations could
be made.  The window function assumes stellar parameters appropriate
for a mid M~dwarf star (\Msun=0.2 \Mstar\ \brk{\Msun} ; \Rstar=0.26
\brk{\Rsun}).}

In addition to the window functions assuming at least $k\geq$ 1, 2, 3,
and 5 transit events are observed (solid lines from top to bottom,
respectively), Figure~\ref{fig:winfunc} also shows the window function
(solid orange line) which corresponds to the triggering mode of operation
adopted by the \Mearth\ project \citep{NUT08}.  Real-time analysis of
the data identifies a potential flux decrement for a target, and the
observing scheduler interrupts cycling through targets and enters a
high cadence follow-up and stare mode in order to get a high
sensitivity detection of the transit egress.  This has the potential
of detecting a planet with a single transit and relaxed photometric
precision, which extends the detectability to longer orbital periods
and enables sampling more targets.  However, in practice for M~dwarfs,
the transit duration $\lesssim 1$ hr and the transit must be observed
during the first portion of the transit event to allow sufficient time
to commence the high cadence staring mode.  Thus, \Tdur\ must be
reduced to account for the triggering mode delay.  Throughout, we
reduce \Tdur\ by $\tau_{\rm trig}$=30 min when simulating the
triggering mode.  For the mid M~dwarf example shown in
Figure~\ref{fig:winfunc}, the triggering mode window function is
superior to the $k=2$ window function for $P > 4$ day.

\section{Radial Velocity Performance}\label{sec:rv}

In this section we summarize our metrics for the
performance of a radial velocity survey.  We follow the procedure by
\citet{CUM04} to determine the limiting semi-amplitude threshold, $K$,
of a radial velocity survey for a given \nobs\ observations, each of which has radial velocity
precision, $\sigma_{rv}$.  We use the performance of
the Lomb-Scargle periodogram for detecting planetary orbits.  For a
given completeness probability, $p_c$, we invert Equation 24 of
\citet{CUM04} to determine the argument of the error function,
\begin{equation}
\Gamma={\rm erf}^{-1}(2p_{c}-1). \label{eq:gamma}
\end{equation}
The error function
argument, $\Gamma$, relates the expectation of the periodogram in the
presence of a signal, 
\begin{equation}
\langle z_{s}\rangle=(\nu/2)(K^2/2{\sigma_{rv}}^2),\label{eq:exps}
\end{equation}
where $\nu=$\nobs$-5$ is the degrees of freedom, to the detection
threshold of the experiment, 
\begin{equation}
z_{d}=(\nu/2)((m/F)^{2/\nu}-1),\label{eq:detthresh}
\end{equation}
where $m$ is the number of independent
frequencies searched and $F$ is the false alarm probability of the
experiment.  The relationship between Equations~\ref{eq:gamma},~\ref{eq:exps},~and~\ref{eq:detthresh} is given by
\begin{equation}
\langle z_{s}\rangle-z_{d}=2\Gamma \sqrt{\langle z_{s}\rangle}.\label{eq:rvall}
\end{equation}
Equations~\ref{eq:gamma}-\ref{eq:rvall} are taken directly from
\citet{CUM04}.  We numerically solve Equation~\ref{eq:rvall} to derive
the limiting $K/\sigma_{rv}$ that a radial velocity achieves for a
given $p_c$ and \nobs.  We adopt parameters following \citet{CUM04}
that the radial velocity survey has a $F=0.001$ and that $m \approx$
\nobs\ \citep{HOR86}.  An additional assumption for the validity of
these results is that the baseline of observations is longer than the
orbital period of the planet.  In this study we ignore the impact of
uncorrected stellar activity for setting a noise floor on radial
velocity precision \citep{SAA97,BOI11,GOM12}, and the additional
observations necessary to refine the orbital parameters with
sufficient precision to make an accurate prediction for the epoch of
transit \citep{KAN09}.  When analyzing the identification of
transiting planets from a radial velocity detected sample of planets
in this paper, we assume a purely geometric prior transit probability
rather than the potentially 10-25\% higher posterior transit
probability for planets first detected with radial velocities
\citep{STE13}.

\section{Discussion}\label{sec:disc}

With the results from the previous sections, we can evaluate a transit
survey designed to detect a transiting planet and quantify the
equivalent radial velocity survey that is expected to have the same
detection capability for the same total amount of observing time.
{\bff With the framework described in this paper, we assume both the
  radial velocity survey and the transit survey achieve
  \nobs\ observations.  We then determine which method using
  \nobs\ (independent of how long it takes to achieve
  \nobs\ observations) achieves a higher survey completeness level for
  the planet of interest.} One opportunity provided by \jwst\ will be
an unprecedented capability for measuring the atmospheric constituents
of potentially habitable planets \citep{VAL05,DEM09}.  Taking
advantage of the \jwst\ opportunity requires identifying the most
viable transiting planets for follow up by pushing the limits of
photometric and radial velocity precision.  We first compare the
performances of the \Mearth\ transiting survey \citep{NUT08,BER13} and
the \HARPS\ M dwarf radial velocity survey \citep{BON11} for detection
of Super-Earth planets orbiting the brightest M~dwarfs.  Then we show
general results for the design of future surveys pushing the detection
limits down to Earth-sized planets.  The steps for the comparison
begin with determining the \nobs\ necessary for a transit survey to
reach a given percent completeness for an inner HZ planet.  We then
calculate the radial velocity precision required to detect the same
planet with the same amount of observing time.

\subsection{Bright M Dwarfs Hosting A Transiting HZ Super-Earth}\label{sec:discsuperearth}

As an example we calculate the performance for a survey modeled after
the \Mearth\ transit survey \citep{NUT08,BER11}.  The \Mearth\ survey
is focused on identifying transiting Super-Earths orbiting the
brightest M~dwarfs employing the triggering mode of detection along
with ex post facto period-folding search for multiple transit events.
Since \Mearth\ is observing the brightest M~dwarfs, which are far
apart on the sky, it typically can observe only a single target per
telescope pointing.  In order to specify \nobs\ required for a transit
survey, we outline the properties of the planets and their stellar
hosts in Table~\ref{tab:results}.  For a given stellar mass, \Mstar,
the stellar radius, \Rstar, stellar surface gravity \Logg, and stellar
effective temperature, \Teff, are estimated by a spline fit to the
calibration of MK spectral types Table~15.7 and 15.8 of \citet{COX00}.
For the given stellar parameters, we determine the orbital period
corresponding to the inner edge of the HZ, $P_{\rm HZ}$, from the
results of \citet{KAS93}.  Empirically for M dwarfs, we find that an
equilibrium temperature $T_{eq}$ equal to 273 K occurs at the inner
edge of the \citet{KAS93} HZ for a planetary bond albedo, $a=0.3$, and
full redistribution of the stellar flux, $f=1$.  Although a warmer
$T_{eq}=288$~K is a better match to the inner HZ edge of \citet{KAS93}
for higher mass stars, \Mstar$>0.4$~$M_{\odot}$, we use the cooler
$T_{eq}=273$~K to define the inner HZ for all stellar types throughout
this work.

With the stellar parameters and $P_{\rm HZ}$ given, the additional
quantities follow ($\tau_{\rm dur}$, transit~probability, $p_{\rm
  tr}$, radial velocity semi-amplitude, $K$, and transit depth,
$\Delta$) for an Earth-mass and Earth-radius analog as well as a
Super-Earth mass and radius analog to GJ~1214 \citep{CHA09,BER11}.
Table~\ref{tab:results} shows $n$ necessary for a transit survey to
achieve a completeness probability $p_c=0.75$ of witnessing a transit
at least once in the triggering mode of operation at $P_{\rm HZ}$.  We
assume \Tobs=200~day or \Tobs=$8\times P$ whichever is greater.
Focusing on the 0.2 $M_{\star}$ star, $n=449$ are required with
sufficient precision to detect the corresponding transit depth,
$\Delta_{Super-\oplus}=9233$ ppm.  If every star has a planet orbiting
with a period of approximately, $P_{\rm HZ}$, then with $p_{\rm
  tr}=0.0179$, $1/p_{\rm tr}=56$ such targets need to be observed.  At
least $n/(p_{\rm tr}p_c)\sim 34000$ observations are required to
expect to find one transiting planet at the inner HZ orbiting a 0.2
$M_{\star}$ M~dwarf.  Cycling between targets with a cadence of 5 min
and 8 hours of observations, reaching 34000 observations requires 355
clear nights.  {\bff The above results are first-order guidance
  assuming unit planet occurrence.  They can be conveniently scaled to
  the measured occurrence rates of planets orbiting M-dwarfs
  \citep{BON11,DRE13,BER13}.}  Fortunately, the targets are independent, so
this amount of observing time can be achieved with multiple telescopes
working in parallel (\Mearth\ employs eight telescopes and is
commissioning eight more in the Southern hemisphere) or observing
fainter stars with multiple targets per field.  Of course, if the
fraction of stars with such a planet is less than unity, the required
observing time will be more.

It is interesting to note that $n$ is not monotonically increasing
with $M_{\star}$ because the increasing HZ period is offset by the
increase in $\tau_{\rm dur}$ with increasing $M_{\star}$.  In
addition, the trigger mode delay becomes an appreciable fraction of
$\tau_{\rm dur}$ for the lowest mass stars.  For
$M_{\star}<0.6M_{\odot}$, $n\sim500$, thus in principle similar number
of observations are needed to find a HZ planet orbiting a late K as
late M star in the triggering mode of operation.
However, the minimum number of clear nights necessary to achieve the
requisite number of observations $n_{ngt}=\tau_{\rm dur}n/\tau_{\rm
  ngt}$, assuming $\tau_{\rm ngt}=8$~hr observing each night is a
monotonically increasing function of $M_{\star}$. In addition, the
depth of the transit for fixed planet size is $\sim$6 times smaller
for late K hosts than late M hosts placing more stringent requirements
on the photometric precision for observations of late K hosts.

For an individual target we determine the radial velocity precision
necessary to reach the same level of survey completeness for making a
Super-Earth \Msini\ detection at $P_{\rm HZ}$ given the same amount of
observing epochs as the transit survey.  In this case the goal of the
radial velocity survey is to make a sufficient number of
\Msini\ detections in order to identify at least one that also
transits.  For the example at 0.2 $M_{\star}$ the circular orbit
semi-amplitude of a Super-Earth at $P_{\rm HZ}$ is $K\sim$5.2~\mps.
To complete the analysis, we must relate $n$ of a transit survey to an
equivalent $n_{obs,\, rv}$ for a radial velocity survey.  The
\Mearth\ survey cycles between targets with a 5 min cadence on average
and employs eight independent telescopes, and the \HARPS\ M~dwarf
survey had a fixed exposure time of 15 min.  Thus, for the Super-Earth
survey comparison, we assume $n=X\, n_{obs,\, rv}$, where $X=24$,
implying 24 observing epochs over independent targets are obtained in
the transit survey for each radial velocity measurement.  In the case
of a transit survey with multiple targets per field, the conversion
factor can be significantly larger \citep[$X$ is $>$4 orders of
  magnitude larger for a survey like \kepler; ][]{BOR10}.  Using the
same completeness probability $p_c=0.75$ for the radial velocity
survey with a false alarm probability of $F=0.001$, $n_{obs,\,
  rv}=n/24$, and $m=n_{obs,\, rv}$, we use Equation~\ref{eq:rvall} to
estimate the limiting semi-amplitude precision ratio,
$K/\sigma_{rv}=2.8$.  Hence, for the 0.2 $M_{\star}$ case,
$\sigma_{rv}=1.85$~\mps.  To summarize, a radial velocity survey will
have reached the same survey completeness probability for a
Super-Earth orbiting at the inner HZ for a late M~dwarf as a simulated
trigger-mode transit survey if it obtains $n_{obs,\, rv}$ with a
precision of $\sigma_{rv}=1.85$~\mps.  If the radial velocity
precision is $\sigma_{rv}>1.85$~\mps, then the radial velocity survey
will not have reached the same level of survey completeness as the
transit survey and is not as efficient for surveying
\Msini\ detections that also transit their host.

For comparison, the mode of the radial velocity precision from the
{\it HARPS} M~dwarf survey sample is approximately equal to
$\sigma_{HARPS}=3.0$~\mps \citep{BON11}.  In the above example of a
Super-Earth at the inner HZ of $M_{\star}=0.2$ \Msun\ host star, the
inequality, $\sigma_{HARPS}\gtrsim\sigma_{rv}$, implies that formally
the efficiency for transiting planet detection for \Mearth\ is
slightly greater than that for transiting planet
detections among the {\it HARPS} M~dwarf survey \Msini\ detections.
The transiting planet yield not only depends upon relative efficiency,
but also total observing time.  The advantage for \Mearth\ may even be
larger given that the \Mearth\ telescopes are dedicated to the
survey whereas \HARPS\ is a shared instrument.

{\bff The above results depend on the definition of the
inner-HZ \citep{SEL07,KOP13,ZSO13}.  To first order, the results of Section~\ref{sec:onetransit}
show that for a transit survey for fixed detection completeness
\nobs$\propto(P/\tau_{\rm dur})\propto P^{2/3}$.  The expected scaling
with $P$ for a radial velocity survey maintaining the same detection
completeness $n_{obs,\, rv}\propto K^{-2}\propto P^{2/3}$ (see
Equation~26 of \citet{CUM04}) is the same.  Numerical calculations
confirm this expected weak dependence on $P$.  For the 0.2 $M_{\star}$ case,
$\sigma_{rv}$=1.9~\mps\ and 1.7~\mps\ for a cooler, $T_{eq}=240$~K, and
hotter, $T_{eq}=300$~K inner HZ definition, respectively.}

\begin{figure}
\includegraphics[trim=0.3in 0.05in 0.35in 0.25in,clip=true,scale=0.6]{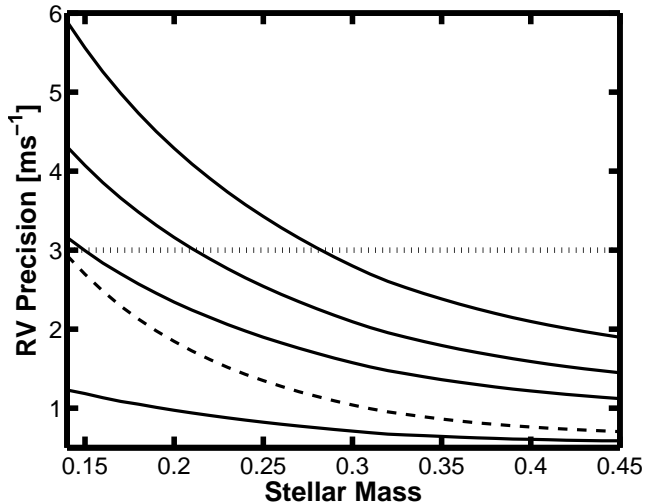}
\caption{
Estimated radial velocity precision ($\sigma_{rv}$) needed to match the efficiency of the 8-telescope \Mearth\ system for finding a transiting inner HZ Super-Earth (GJ 1214 analog) as a function of stellar mass with at least one, two, three, and five transit events required for detection (solid lines), from bottom to top, respectively.  ({\it Dashed line}) shows results for the single transit triggering mode of detection in a transit survey, and the horizontal dotted line indicates the mode of the radial velocity precision achieved by the \HARPS\ M~dwarf survey \citep{BON11}.  The calculation assumes $p_{c}=0.75$ and $X=24$.  \label{fig:rvprecse}}
\end{figure}

The discovery of GJ~1214 did not rely on the triggering mode for
detection, but was detected after phasing two potential transit events
\citep{CHA09}.  Figure~\ref{fig:rvprecse} summarizes the radial
velocity precision required in order to reach the same level of
completeness for a single target using the same amount of observing
time for the transit survey design of requiring $k\geq$ 1, 2, 3, and 5 events
as the solid lines from bottom to top, respectively.  The dash line
shows results for the triggering mode of operation.  The achieved
precision of the {\it HARPS} M-dwarf survey $\sigma_{HARPS}=3.0$ is
also shown (horizontal dotted line) demarcates the stellar hosts
where a 8-telescope system like \Mearth\ is more (below) and less
(above) efficient.  In the triggering discovery mode (dashed line),
\Mstar$>$0.15\Msun\ is the stellar host regime where the
8-telescope \Mearth\ system in the triggering discovery mode is more
efficient than \HARPS\ for detecting transiting Super-Earth planets.
For a 3-transit requirement detection mode, the detection efficiency
equivalency is at \Mstar$\sim$0.22\Msun. 

\begin{figure}
\includegraphics[trim=0.2in 0.05in 0.35in 0.2in,clip=true,scale=0.6]{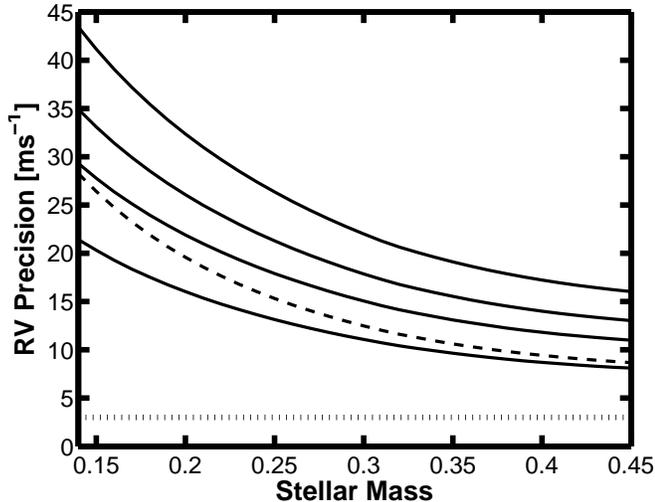}
\caption{
Same as Figure~\ref{fig:rvprecse}, but for a single ${\rm M sini}=6.6$
\Mear\ planet detection that does not necessarily
transit.\label{fig:rvpreconedetse}}
\end{figure}

The results shown in Figure~\ref{fig:rvprecse} and the previous
discussion assume the goal of the survey is to find a HZ planet that
actually transits.  Given the small probability to transit, the radial
velocity survey will achieve a large number of detections $\propto
1/p_{\rm tr}$ for every single HZ planet that transits.  If the goal
of the radial velocity survey is just to detect a single
\Msini\ Super-Earth mass object, then it is more appropriate to
compare to the time necessary for a transit survey to result in a
single detection, $n_{obs,\, tr}/p_{\rm tr}$.  The requisite radial
velocity precision for a single radial velocity detection of a
Super-Earth with the equivalent observing time for a single transiting
planet detection using the 8-telescope \Mearth\ system is shown in
Figure~\ref{fig:rvpreconedetse}.  For measuring the occurrence rate of
Super-Earth \Msini\ planets \citep{BON11}, the \HARPS\ M~dwarf survey
is predicted to have superior efficiency for measuring Super-Earth
population statistics compared to the \Mearth\ survey.  However, a
transit survey observing fainter targets can achieve a much larger
conversion factor, $X$.  A transit survey requiring $k\geq3$ transits
for detection with a multi-plexing conversion factor of $X\sim2200$
has equivalent detection efficiency to a radial velocity survey with
$\sigma_{rv}=2.0$~\mps\ for inner HZ Super-Earth planets around late M
dwarfs.

One caveat to the discussion is the need for $1/p_{\rm tr}>56$
\Msini\ detections at the inner HZ for late M dwarfs before one
expects to find one that also transits.  Both the \HARPS\ and
\Mearth\ surveys target the brightest M dwarfs available, but due to
the geometric transit alignment one must survey a larger volume and
fainter magnitudes for the \Msini\ detections that also transit.  For
uniform stellar density, the brightest transiting planet at the inner
HZ for late M dwarfs will typically be $\Delta_{mag}\sim 3.0$ mag
fainter than the brightest \Msini\ nontransiting detection.  At the
median brightness of the \Mearth\ M~dwarf sample (V=15.2, J.\ Irwin,
{\it private communication}) the \HARPS\ instrumental radial velocity
precision is predicted to be $\sigma_{rv}\sim 10$~\mps \citep{BON11}.
Thus at the median brightness of the \Mearth\ M~dwarf sample,
\Mearth\ is much more efficient at finding transiting Super-Earth
planets at the inner HZ than if attempting to survey the
\Mearth\ M~dwarf sample for transiting planets with \HARPS.

\subsection{Bright M Dwarfs Hosting A Transiting HZ Earth Analog}\label{sec:discearth}

We repeat the analysis of the previous section down to the regime of
truly Earth-size and Earth-mass planets at the inner HZ of the
brightest M~dwarfs.  The {\it TESS} mission \citep{RIC10} is one
possibility for identifying transiting HZ Earth analogs orbiting
bright stars, but here we briefly explore the possibility of a
scaled-up \Mearth\ survey and the associated equivalent radial
velocity precision.

From the perspective of window functions, the required $n_{obs,\, tr}$
is the same as for the detection of Super-Earths at the inner HZ.
However, the photometric precision must be improved by a factor
$\sim$7 from the Super-Earth survey to detect the shallower
Earth-analog transit depth.  In principle, the requisite photometric
precision for enabling the trigger mode of detection for Earth-size
planets orbiting M~dwarfs has been demonstrated from the ground.
Scintillation limited performance using 4m class telescopes has
achieved $\sigma_{p}$=250 ppm min$^{-1}$ photometric noise rate on
V$\sim$12 for fields with a comparable brightness and spectral type
companion in the optical \citep{GIL93,BUR10}.  \citet{MAN11} achieve
1250 ppm min$^{-1}$ photometric noise rate on V$\sim$13 late K and
early M dwarfs with a 2m class telescope with an innovative target
cycling approach.  Thus, the practical limitation of a scaled up
\Mearth\ survey is available observing time on larger telescopes
rather than a physical limit imposed by the Earth's atmosphere.

\begin{figure}
\includegraphics[trim=0.2in 0.05in 0.35in 0.2in,clip=true,scale=0.6]{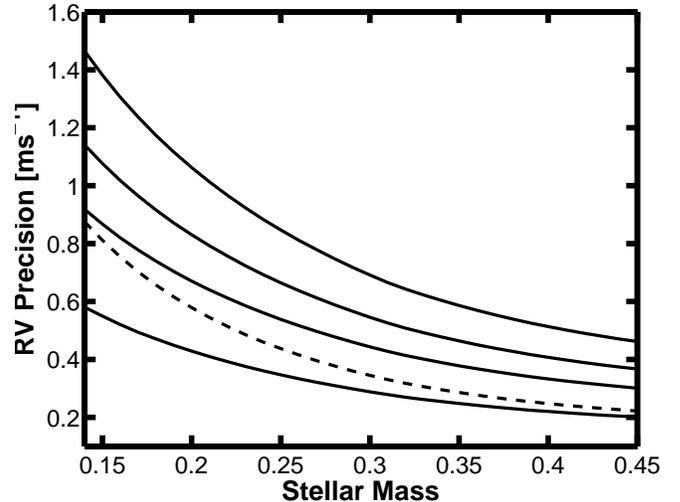}
\caption{
Radial velocity precision needed to match the efficiency of a single-target transit survey for finding a transiting inner HZ Earth analog as a function of \Mstar\ where the transit survey requires at least one, one with triggering (dotted line), three, and five transit events for detection, bottom to top, respectively.   The calculation assumes $p_{c}=0.75$ and $X=10$. \label{fig:rvprec}}
\end{figure}

\begin{figure}
\includegraphics[trim=0.2in 0.05in 0.35in 0.2in,clip=true,scale=0.6]{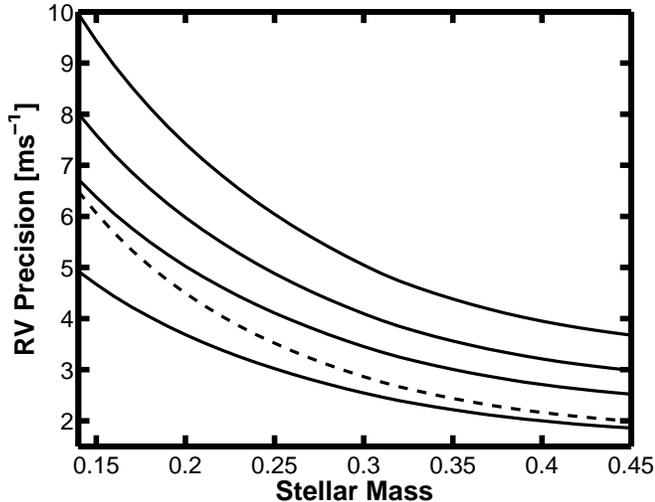}
\caption{
Same as Figure~\ref{fig:rvprec}, but for a single ${\rm M sini}=1$ \Mear\ planet in the inner HZ that does not necessarily transit.
Note the change in vertical scale compared to Figure~\ref{fig:rvprec}. \label{fig:rvpreconedet}}
\end{figure}

Carrying through the above calculation for an Earth analog at the
inner HZ, we find $\sigma_{rv}=0.58$~\mps\ is needed for comparable
efficiency to the triggering mode transit survey.  For this result we
assume $X=10$, which would be expected given the efficiency difference
between a high resolution spectrograph for precision radial velocity
measurements compared to single-target photometry.  Since radial
velocity instrumentation is typically on larger aperture
telescopes\footnote{The~multi-telescope~radial~velocity~concept~\citep{BOT13}~is~another~possibility -- http://www.astro.caltech.edu/minerva/Home.html},
assuming more efficient radial velocity observations, $X=5$, results in
$\sigma_{rv}=0.87$ ms$^{-1}$.  The noise floor for M~dwarf precision
radial velocities was $\sigma_{rv}=1.1$~\mps\ for the \HARPS\ M~dwarf
survey \citep{BON11}.  Results for additional transit survey designs
of 2, 3, or 5 transits for detection are shown in
Figure~\ref{fig:rvprec}, and the precision needed to reach comparable
efficiency of an \Msini\ detection is shown in
Figure~\ref{fig:rvpreconedet}.

\section{Conclusion}\label{sec:conclusion}

Pushing the ground-based photometric and radial velocity techniques
provides an opportunity for detecting true Earth-sized planets at the
inner HZ orbiting the brightest late M~dwarfs.  For example a scaled
up \Mearth\ survey could detect HZ Earth-sized transiting planets
among the brightest late M~dwarfs.  We predict that for the same
amount of observing time, a radial velocity survey requires precision
$\sigma_{rv}<0.6$~\mps\ in order to more efficiently detect a \Msini=1
\Mear\ transiting planet at the inner HZ for mid M~dwarf.  From space,
the {\it TESS} mission \citep{RIC10} plans to survey early M and K
dwarfs for such Earth analogs.  There is great impetus for finding the
rare Earth analogs that also transit among the brightest stars to take
full advantage of the follow-up potential of \hst\ and the upcoming
opportunities provided by \jwst.  \jwst\ will usher in an unprecedented
capability for measuring the atmospheric constituents of potentially
habitable planets \citep{VAL05,DEM09}.

We demonstrate that the window function for a transit survey can be
expressed in terms of the hypergeometric probability distribution.
An analytic approximation to the window function simplifies the design of
a transit survey and provides an accurate estimate to the observing
time necessary to achieve a given level of completeness for detection
of a transiting planet.  {\bff Furthermore, to first order, the
  resulting window function for a transit survey depends only on the
  total number of independent observations and secondarily depends
  upon the distribution of observation times or baseline over which
  the observations are obtained.
We employ these results regarding the transit survey window function model in order to determine the comparable radial velocity survey
precision that achieves the same completeness level given the same
number of observations}.  For the brightest M~dwarfs, $V<12$, we
predict \HARPS\ and the \Mearth\ surveys have approximately the same
efficiency for detecting Super-Earth planets that transit and orbit at
the inner HZ.  However, at the median brightness of the
\Mearth\ sample, $V=15$, the \Mearth\ survey is predicted to have much
higher efficiency at detecting transiting Super-Earths than
\HARPS\ due to the degraded radial velocity precision for fainter
stars.  The higher efficiency is predominately driven by the small
geometric transit probability at the inner HZ for late M~dwarfs.
\HARPS\ is predicted to have much higher efficiency in
\Msini\ detections than \Mearth\ has for transiting planets, thus
making \HARPS\ efficient at determining Super-Earth \Msini\ population
statistics.  However, complementary planet population studies are
possible from wide field transit surveys that observe many targets per
observing epoch \citep[$X\sim10^{5}$ in the case of \kepler; ][]{YOU11,HOW12,DON13,FRE13,DRE13,PET13}.

\section{APPENDIX: Analytic Approximations to the Window Function}\label{ref:apx}

\subsection{Binomial}

Prior work \citep[e.g. Eq. A2 of][]{DEE04} adopted approach A of Section~\ref{sec:transit},
 ``each time the planet transits, were we observing?''
{\bff For this statistical model of the window function each orbital period
time interval represents an independent trial.}  Specifically, the
orbital period time interval represents a binomial trial with
probability of our witnessing it equal to the average fraction of time
spent observing $f_{\rm cov}$.  Thus, the probability of witnessing
exactly $k$ transits is
\begin{equation}
p_k = {M \choose k}f_{\rm cov}^k (1-f_{\rm cov})^{M-k}, k = 0, 1, 2, ..., M \label{eq:binomial}
\end{equation}
recalling that the expectation value for the number of trials $M=$\Tobs$/P$ (Eq. \ref{eq:M}) and that
the probability of witnessing $k$ or more transits $p_{\geq k}$ is given by Eq. \ref{eq:probN}.

\subsection{Poisson} \label{sec:poisson}

{\bff Approach A also can be addressed, in appropriate limits, specifically large M and small $f_{\rm cov}$} (e.g. Appendix 4 of Lyons 1996), using the Poisson approximation to the binomial distribution, 
\begin{equation}
p_k = {\lambda^k\over{k!}} e^{-\lambda}, k = 0, 1, 2, ..., M \label{eq:poisson}
\end{equation}
where $\lambda = M f_{\rm cov}$.
Even though the Poisson distribution is defined for all $k \ge 0$, the maximum number of
transits in time span \Tobs\ is M = $\lfloor$\Tobs/$P\rfloor+1$. The physical interpretation of the Poisson process
would be that transits are observed at random times at an average rate of $\lambda$ transits in \Tobs, which approaches $f_{\rm cov}/P$ in the
appropriate limit of large $M$. Of course, transits do not occur at random times, but the approximation
works if observations are sparsely and randomly distributed throughout the time span \Tobs.
\citet{DEE04} recommend that the Poisson approximation gives good results so long as $f_{\rm cov} \la 0.5$ and $M$ is large.  {\bff While it is the least accurate, the Poisson approximation is the simplest to compute; we recommend it for hand calculations (see Section~\ref{sec:onetransit}) and to validate more complicated window function implementations.}

\subsection{Hypergeometric} \label{sec:hyperg}

\begin{figure}
\includegraphics[trim=0.3in 0.2in 0.4in 0.0in,clip=true,scale=0.5]{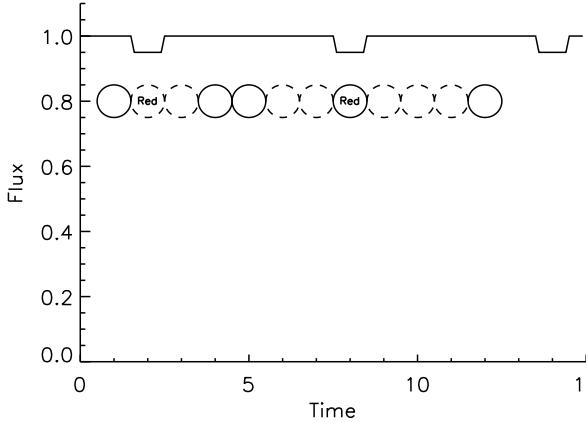}
\caption{
Planetary transit observability can be modeled as drawing balls
without replacement. A schematic of a transit light curve, with $P=6$ units and with depth
and width exaggerated for clarity, is above a set of 
colored balls.  Each ball in the contiguous line of balls corresponds to a
time interval equal to the transit duration.  
Between the first and last observations, there
are $N=12$ time intervals total: \nobs$=5$ intervals (solid circles) each with
sufficient observational sensitivity to detect a transit if it had
occurred during the given interval, and seven intervals (dashed
circles) each with insufficient or no observations during the given
interval. 
Red balls (labeled) correspond to $M=2$ intervals in which a transit occurs; all
other balls are white (no label). 
\label{fig:cartoon}
}
\end{figure}

{\bff In approach B, ``each time we observe, is the planet transiting?'',
the statistical model of the window function adopts the transit
duration time interval as the relevant independent trial.} We make an
analogy between searching for a transiting planet and randomly drawing
colored balls from an urn without replacement
(Figure~\ref{fig:cartoon}).  Each ball represents a time interval
equal to the transit duration \Tdur\ and also a potential observation
of a transit.  The parameter \nobs\ in the hypergeometric distribution
is the number of unique observation intervals of length \Tdur\ that
have sufficient sensitivity to detect a transit if it had occurred
during the interval.  Parceling the observing span \Tobs\ into
intervals of \Tdur\ results in a set of $N=$\Tobs/\Tdur\ time
intervals, or by analogy $N$ balls in the urn.  In the illustration,
the orbital period $P$ is six times the transit duration \Tdur. In
general, for circular orbits and central transits, the ratio
$P$/\Tdur$\approx \pi a/R_{\star}$, where $a$ is the orbital semimajor
axis and the formula neglects the planetary radius.  The illustration
shows a contiguous set of $N=$\Tobs/\Tdur$=12$ balls, where $M=2$
balls are red (in transit), and the $N-M=10$ remaining balls are
colored white (out-of-transit).  There are \nobs=5 balls drawn with
solid outlines (observations) and $N-$\nobs$=7$ balls with dashed
outlines (observation gaps or observation intervals with insufficient
sensitivity to detect a transit even if it had occurred in the
particular interval).  Thus the the probability of having $k$
observations occur during transit among the \nobs\ observations is
given by the hypergeometric distribution,
\begin{equation}
p_k={ { {{M\choose k} { {N-M}\choose{n-k}} } }  \over  {N\choose n} }, k = 0, 1, 2, ..., M \label{eq:hyperg}
\end{equation}
which is appropriate for this case of sampling without replacement. In the nomenclature
of statistics,
\begin{eqnarray} 
M &=& {\rm sub~population~size,}\\
N &=& {\rm total~population~size, and} \\
n &=& {\rm sample~size.}
\end{eqnarray}

A limitation common to the binomial and hypergeometric distributions is that their
input parameters formally are required to be integers, because of the combinatorics
operator ``choose.'' For example, each equation contains $M$ choose $k$; that $k$ is an
integer is not a problem, but if $M$ is forced to also, then the window function can have
a stair-step appearance. This is readily mitigated by using implementations of the
operators that can handle real-valued inputs via the gamma function, e.g.
\begin{equation}
{M\choose k} = { {M!}\over{ {k! (M-k)!} } }
 = { {\Gamma(M+1)}\over{ {\Gamma(k+1)\Gamma(M-k+1) } } }
= { {M}\over{ {k(M-k)} } }
{ {\Gamma(M)}\over{ {\Gamma(k) \Gamma(M-k)} } } . \label{eq:mck}
\end{equation}
One may wish to use Stirling's formula or a similar approximation for the factorials, or
instead of the gamma function itself, one may wish to use a log(gamma) function if available, to
avoid numerical overflow.

It is possible to show the numerical similarities of the binomial
model of \citet{DEE04} with the hypergeometric model
of this study by expressing $n=f_{\rm cov}T_{\rm obs}/\tau_{\rm dur}$,
then the binomial model can be written as $B(k;M,f_{\rm
  cov})=B(k;T_{\rm obs}/P,n\tau_{\rm dur}/T_{\rm obs})$.  The
expectation of the binomial model of \citet{DEE04}, $E[B(M,f_{\rm
    cov})]=f_{\rm cov}M=f_{\rm cov}T_{\rm obs}/P=n\tau_{\rm dur}/P$, is the same as the
expectation of the hypergeometric model of this study,
$E[H(N,M,n)]=nM/N=n\tau_{\rm dur}/P$.  The approaches' higher moments are not identical; For example, the
variance ratio between the models, $Var[B(M,f_{\rm
    cov})]/Var[H(N,M,n)]=(1-\tau_{\rm dur}/T_{\rm obs})/(1-\tau_{\rm
  dur}/P)$.  However, since $\tau_{\rm dur}\ll P<T_{\rm obs}$, for
parameters generally applicable to a transit survey, the variances are nearly equal.

\subsection{Probability of Witnessing at least one Transit or two Transits}\label{sec:onetransit}

The probability of witnessing at least one transit (from Equation~\ref{eq:probN}) is
\begin{eqnarray} 
p_{\geq 1} &=& e^{-\lambda}  ~~~~~~ ~~~~~~  ~~~~~~  ~~~~~~ ~~~~~~  ~~~~~~  ~~~~~~   {\rm Poisson,}\label{eq:poione} \\
  &=& 1 - (1-f_{cov})^M   ~~~~~~ ~~~~~~  ~~~~~~ ~~~~~~  {\rm binomial, or} \\
  &=& 1 - { { {N-M}\choose n}\over{N\choose n} } ~~~~~~ ~~~~~~ ~~~~~~ ~~~~~~ {\rm hypergeometric.}
\end{eqnarray}
Associating $f_{\rm cov}=n\tau_{\rm dur}/T_{\rm obs}$, Equation~\ref{eq:poione}
is readily inverted: \nobs$=-\ln(1-p_c)(P/\tau_{\rm dur})$.
This provides a convenient benchmark for planning purposes as well
checking how probable a potential planetary discovery is in a transit
search.

The probability of witnessing at least two transits is
\begin{eqnarray} 
p_{\geq 2} &=& e^{-\lambda} (1 - \lambda)  ~~~~~~ ~~~~~~ ~~~~~~ ~~~~~~ ~~~~~~  ~~~~~~ ~~~~~~ {\rm Poisson,} \\
  &=& 1 - (1-f_{cov})^M - M f_{cov} (1-f_{cov})^{M-1}  ~~~~~~ {\rm binomial, or} \\
  &=& 1 - { { {N-M}\choose n}\over{N\choose n} } - M { { {N-M}\choose {n-1}}\over{N\choose n} } ~~~~~~ ~~~~~~ ~~~~~~  ~~~~~~    {\rm hypergeometric.}
\end{eqnarray}

\subsection{Comparisons to Numerical Simulations}\label{sec:numericalcomp}

In Figure~\ref{fig:beattycomp} we compare the resulting window
function given by Equation~\ref{eq:probN} to the exact numerical
window function for a hypothetical ground-based transit survey that
observes for 30 consecutive days with observations for 8 hours each
night.  The specified transit survey is the same as presented in
Figure~1 of \citet{BEA08} to demonstrate their analytical
representation of a transit survey (solid blue line).  The exact
numerical window function (dash black line) shows the substantial
alias structure that is typical of ground-based transit surveys due to
the day-night observing cycle, and which is not modeled by the simpler
analytic approximations.

The analytic expressions approximate well the average window function.
To demonstrate this we show in Figure~\ref{fig:beattycomp} a moving
average of the exact numerical function using a 1~day window (solid
black line).  The left and right panels provide results for requiring
$k\geq 2$ and $k\geq 3$ transit events for detection, respectively.
We show the hypergeometric window function model from this study
(Equation~\ref{eq:hyperg}) as the solid red line.  Solar values for
the stellar parameters are assumed.  The Poisson window function model
(Equation~\ref{eq:poisson}, dash red line) also does a reasonable job
in this scenario with $f_{\rm cov}=0.33$.

Figure~\ref{fig:winfunc} shows the window function for a transit
survey that operates in a low duty cycle mode obtaining several
observations per night over the course of an observing season.  We
simulate this mode of operation through specifying a time series with
\nobs$=200$ that is expected for a target from a single observatory
over the course of an observing season including losses due to weather
and stellar parameters appropriate for a mid M~dwarf star.  The exact
numerical window functions assuming at least $k\geq$ 1, 2, 3, and 5
transit events are observed (solid lines from top to bottom,
respectively).  The hypergeometric model (solid red line) and its
Poisson model approximation (dash blue line) agree extremely well as
expected for a low duty cycle transit survey.

\begin{figure}
\includegraphics[trim=0.3in 0.15in 0.4in 0.0in,clip=true,scale=0.5]{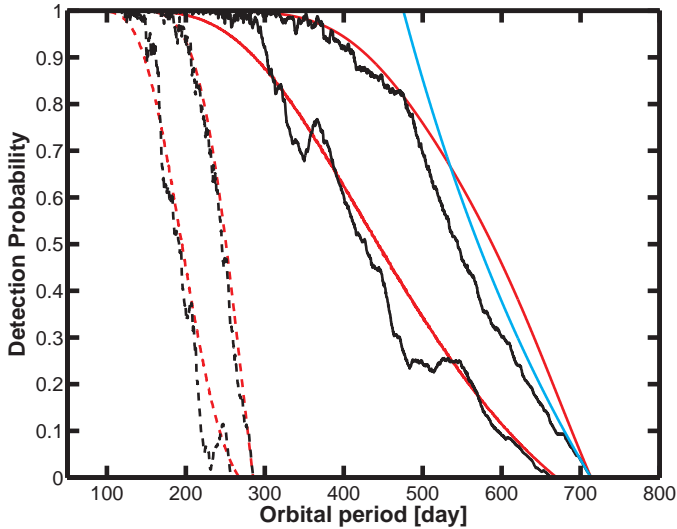}
\caption{
Detection probability (upper solid black line) for a representative target observed by \kepler\ for 16 quarters and requiring $k\geq3$ transits for detection.  Detection probability (lower solid black line) for a \kepler\ target impacted by yearly data gaps of one missing quarter per year.  Analytic window functions are shown using the hypergeometric approximation (Equations~\ref{eq:probN}~and~\ref{eq:hyperg}) for three or more transits, $k\geq 3$ (solid red lines).  Also shown is the equivalent set of lines as above but requiring $k\geq6$ transits for detection (dash lines) and the analytical window function for continuous observations with no gaps from \citet{FLE08} (blue solid line).
\label{fig:keplerwin}}
\end{figure}

Figure~\ref{fig:keplerwin} shows the window function for two typical
targets observed by \kepler\ using Quarters 1 through 16 ($\sim$3.6
yr) of nearly continuous coverage \citep{TEN13}.  The rightmost set of
black solid, red solid, and blue solid lines correspond to the
coverage of \kepler-11, host to six transiting planets \citep{LIS11}.
This target was observed for all 16 \kepler\ Quarters.  For the time
series, we employ barycentric julian date mid-cadence times stamps
that have valid data as determined by the \kepler\ pipeline
\citep{SMI12,WU10}.  The hypergeometric model (solid red line)
requiring $k\geq$3 transit events being observed agrees well with the
exact numerical calculation (solid black line) also requiring
$k\geq3$.  As expected the analytical window model function for
continuous observations with no gaps of \citep[Table~2 of ][]{FLE08},
overestimates the window function due to the reality of gaps in
\kepler\ data.  The leftmost set of solid black and solid red lines
correspond to the coverage of the host to {\it TrES}-2b
\citep[a.k.a.\ Kepler-1b; ][]{ODO06B,BAR12}.  This target is impacted
by the loss of a detector in Quarter four \citep{BAT13}, causing missing
data every fourth \kepler\ Quarter.  The detector loss impacts
$\sim$19\% of \kepler\ targets having reduced amount of data
available.  The dash lines are the same as above, but they require
$k\geq6$ transits for detection.

The hypergeometric model does not fully capture the details present in
the exact numerical window functions.  The assumptions of the
hypergeometric model begin to break down in observing scenarios
consisting of high duty cycle blocks interspersed with long periods of
no data.  In addition to the sharp, isolated aliases at the diurnal
cycle and its harmonics prevalent in ground-based transit survey
window functions (which we readily average over), there are
``broad-band'' alias structures present across all orbital periods
that can result in the exact numerical window function to
deviate from the hypergeometric model.

In Figure~\ref{fig:alternatobs} we demonstrate the principal
dependence of a transit survey window function on $n$ and the weak
dependence on \Tobs.  We show the hypergeometric model (solid lines)
and the moving average of the exact numerical window function (dash
lines) for three scenarios that have the same $n$ but vary in \Tobs.
The left and right panels vary by requiring $k\geq$ 2 and 3 transit
events for detection, respectively.  The baseline transit survey
consists of 7 hours of observations each day for 30 consecutive days
(black lines).  We simulate a longitudinally-distributed network
survey by considering a transit survey with half \Tobs\ (15
consecutive days) of the baseline case while doubling the observing
time each day to conserve $n$ (red lines).  The window function does
show an increase in detection probability at intermediate periods at
the cost of removing the low detection probability tail toward longer
periods of the baseline scenario.  The third $n$-conserving scenario
consists of splitting the 30 consecutive days of observations from the
baseline case into two 15 consecutive day groups separated by a year
(blue lines) giving a \Tobs=380 day.  In contrast to the
$n$-conserving cases, we show a transit survey window function
resulting from doubling $n$ while keeping \Tobs\ the same as the
baseline case (magenta lines).  These scenarios demonstrate that the
detection probability of a transit survey are very sensitive to $n$.

\begin{deluxetable}{@{\hspace{2 pt}}c@{\hspace{2 pt}}c@{\hspace{2 pt}}c@{\hspace{2 pt}}c@{\hspace{2 pt}}c@{\hspace{2 pt}}c@{\hspace{2 pt}}c@{\hspace{2 pt}}c@{\hspace{2 pt}}c@{\hspace{2 pt}}c@{\hspace{2 pt}}c@{\hspace{2 pt}}c@{\hspace{2 pt}}c@{\hspace{2 pt}}c@{\hspace{2 pt}}c@{\hspace{2 pt}}}
\tabletypesize{\small}
\tablewidth{0pt}
\tablecaption{{\rm Stellar, Transit, \& Radial Velocity Parameters}}
\startdata
\hline
\hline
\Mstar\ & \Rstar\  & \Teff\  & $P_{\rm HZ}$  & $p_{\rm tr}$ & $\Delta_{\oplus}$  & $\Delta_{{\rm S}\oplus}$ & $\tau_{\rm dur}$ & \nobs\tablenotemark{a} & ${\rm n_{ngt}}$ & \nobs$/p_{\rm tr}$ & $K_{{\rm S}\oplus}$  & $\sigma_{\rm rv\, S\oplus}$\tablenotemark{b} & $K_{\oplus}$  &  $\sigma_{\rm rv\, \oplus}$\tablenotemark{c}\\ 
 \brk{\Msun} & \brk{\Rsun} & \brk{K} & \brk{day} &  &   &  \brk{ppm}       &  \brk{ppm}  &  \brk{hr}  &  &  &  \brk{\mps} &  \brk{\mps} & \brk{\mps} &  \brk{\mps}\\
\hline
0.15 & 0.20 & 3077 &  10.47 & 0.0188 & 2167 & 15801 &  1.18 &  492 &   73 &  26139 & 6.83 &  2.69 & 1.035 & 0.81  \\
0.17 & 0.22 & 3109 &  11.62 & 0.0184 & 1724 & 12568 &  1.29 &  472 &   76 &  25606 & 6.07 &  2.29 & 0.920 & 0.70 \\
\tablenotemark{d}{\bf 0.20} & {\bf 0.26} & {\bf 3155} &  {\bf 13.61} & {\bf 0.0179} & {\bf 1266} & {\bf 9233} &  {\bf 1.46} &  {\bf 449} &  {\bf 82} &  {\bf 25086} & {\bf 5.17} &  {\bf 1.85}  & {\bf 0.783} & {\bf 0.58}\\
0.22 & 0.28 & 3185 &  15.13 & 0.0176 & 1052 & 7671 &  1.59 &  436 &   87 &  24825 & 4.68 &  1.62  & 0.709 & 0.51\\
0.25 & 0.32 & 3230 &  17.75 & 0.0171 & 818  & 5967 &  1.82 &  421 &   96 &  24655 & 4.08 &  1.35  & 0.618 & 0.44\\
0.27 & 0.35 & 3261 &  19.75 & 0.0168 & 703  & 5127 &  1.99 &  413 &  102 &  24653 & 3.74 &  1.21  & 0.566 & 0.40\\
0.30 & 0.38 & 3310 &  23.14 & 0.0163 & 572  & 4171 &  2.26 &  404 &  114 &  24849 & 3.30 &  1.04  & 0.501 & 0.34\\
0.35 & 0.44 & 3404 &  29.86 & 0.0154 & 426  & 3110 &  2.75 &  404 &  139 &  26278 & 2.74 &  0.86  & 0.415 & 0.28\\
0.40 & 0.50 & 3520 &  37.76 & 0.0144 & 336  & 2449 &  3.26 &  418 &  170 &  29071 & 2.32 &  0.76  & 0.351 & 0.25\\
0.45 & 0.55 & 3661 &  46.32 & 0.0133 & 278  & 2025 &  3.69 &  442 &  204 &  33257 & 2.00 &  0.70  & 0.303 & 0.22\\
0.50 & 0.59 & 3811 &  55.12 & 0.0123 & 239  & 1745 &  4.06 &  473 &  240 &  38559 & 1.76 &  0.67  & 0.267 & 0.20\\
0.55 & 0.63 & 3952 &  64.07 & 0.0114 & 213  & 1555 &  4.38 &  503 &  276 &  44098 & 1.57 &  0.63  & 0.238 & 0.19\\
0.60 & 0.66 & 4105 &  74.87 & 0.0106 & 192  & 1402 &  4.75 &  538 &  319 &  50889 & 1.41 &  0.61  & 0.213 & 0.18\\
0.70 & 0.75 & 4591 & 117.39 & 0.0085 & 148  & 1078 &  5.95 &  657 &  489 &  77718 & 1.09 &  0.56  & 0.166 & 0.15\\
0.80 & 0.86 & 5198 & 189.37 & 0.0066 & 114  &  834 &  7.49 &  827 &  775 & 125410 & 0.85 &  0.53  & 0.129 & 0.14\\
0.90 & 0.90 & 5509 & 231.34 & 0.0059 & 103  &  749 &  8.15 &  924 &  941 & 157421 & 0.73 &  0.49  & 0.112 & 0.13\\
1.00 & 1.03 & 5794 & 307.04 & 0.0053 & 80   &  581 &  9.78 & 1011 & 1235 & 190520 & 0.63 &  0.44  & 0.095 & 0.11\\
1.10 & 1.16 & 6063 & 404.58 & 0.0048 & 63   &  459 & 11.76 & 1097 & 1613 & 226371 & 0.54 &  0.40  & 0.081 & 0.10\\
\hline
\enddata
\label{tab:results}
\tablenotetext{a}{Single transit triggering mode of operation and $p_c=0.75$ completeness level.}
\tablenotetext{b}{Assumes $X=24$ (see Section~\ref{sec:discsuperearth}).}
\tablenotetext{c}{Assumes $X=10$ (see Section~\ref{sec:discearth}).}
\tablenotetext{d}{The bold face row is discussed in Section~\ref{sec:discsuperearth}.}
\end{deluxetable}

\acknowledgments

This work benefited from discussions with Cullen Blake, B.~Scott Gaudi, Jonathan Irwin, Jon Jenkins, and George Ricker.  We thank the referee for insightful suggestions which improved the manuscript.

\end{document}